\input epsf
\documentstyle[12pt]{article}

\def\suthree{$\rm{SU(3)}_L \! \! \times \!\rm{SU(3)}_R$}

\def\sl{\hspace{-.55em} \slash}
\def\half{{1 \over 2}}
\def\simgeq{\: \raisebox{-.8ex}{$\stackrel{\textstyle{>}}{\sim}$} \:}

\title{Strange Baryonic Matter from Chiral Effective Lagrangians}
\author{Bruce Ritzi and Graciela Gelmini   
\\Dept. of Physics, University of California, Los Angeles
\\Los Angeles, California 90095}
\date{October 1996} 

\begin{document}

\maketitle

\begin{abstract}
We investigate the existence of bound states of baryons
in a kaon condensate using chiral mean field theory.
 The interactions are described by an effective \suthree\
chiral lagrangian where terms of higher order in density, baryon momentum,
and kaon mass are suppressed by powers of 
the symmetry breaking scale, $\Lambda$. We take up
to next to leading order terms ($n = 2,3,4$).
We search for infinite baryon number solutions, namely ``strange
baryonic matter'',
using a Thomas-Fermi
approximation for a slowly varying condensate and a lowest order
Hartree approximation to describe the many body interactions.
For simplicity we study a pure $K^0$ condensate and only neutrons,
the lightest baryons in that condensate. We find solutions with
neutron number densities, $\rho_n \simgeq 3.5 \rho_0$, where
$\rho_0$ is the infinite nuclear matter density. This is
consistent with the estimate of the onset of a K-condensate at
$\rho_n \simeq$ 2--4 $\rho_0$. We show that the binding energies, $E_b$,
grow with $\rho_n$ and for $\rho_n < 7 \rho_0$ (at $\rho_n
\simgeq 7 \rho_0$ perturbative expansion is lost) we find
$E_b < 150 MeV$ ($E_b < 70 MeV$ for $\rho_n < 5 \rho_0$) even in the
most favorable cases. These binding energies may be too low
for this type of matter to appear and persist in the early
universe.  

\end{abstract}
\vspace{-7in}
hep-ph/9610496
\begin{flushright}
\vspace{-1.2cm}
UCLA/96/TEP/25 \\
October, 1996
\end{flushright}
\pagebreak

\begin{center}
\section*{\normalsize \bfseries{1. Introduction}}
\end{center}
\setcounter{section}{1}

QCD is the theory of strong interactions among quarks and gluons,
the elementary particles that constitute hadrons. With three
massless quarks, $m_u = m_d = m_s = 0$, the QCD lagrangian has a 
global chiral \suthree symmetry. This is considered to be an 
accidental symmetry, since there is no deeper reason for the
(almost) masslessness of these three quarks. The chiral symmetry is 
approximate since the three quarks masses are small, but not zero.
The symmetry is most explicitly violated by the s-quark mass,
$m_s \simeq 240MeV$, which is much larger than the u and d masses
($m_u \simeq 6~MeV, m_d \simeq 12~MeV$). Even if we can not derive
from QCD the properties of the quark and gluon bound states,
the hadrons, we know the mass spectrum and low momentum 
interactions of hadrons show the persistence of the \suthree
chiral symmetry. The $SU(3)_V$ vectorial and $SU(3)_A$ axial vector symmetry
groups (those whose generators are the sum and the difference
respectively of the left and right generators) are realized
differently. The vectorial subgroup $SU(3)_V$ is realized
in the Wigner--Weyl mode, yielding (almost) degenerate
multiplets (octets and tenplets) of baryons and mesons.
This is the (approximate) symmetry used to classify 
hadrons (in the ``eightfold way''), that lead to the proposal
of quarks as a means to populate the fundamental representation of
the group. The axial symmetry $SU(3)_A$ is instead
realized in the Nambu--Goldstone mode, namely it is spontaneously
broken at a scale $\Lambda \simeq 1GeV$ yielding an octet 
of (quasi) Goldstone bosons, one for each broken
generator, the pions, kaons, and eta mesons. The lightness of
these mesons compared to the other hadrons justifies this 
identification. The mass of the Goldstone bosons is a result
of the explicit breaking of $SU(3)_A$ due to the non-zero
quark masses. In fact from the lowest order ($n=2$, see below)
chiral lagrangian, the $\pi$, $K$, and $\eta$ masses result
linearly proportional to the $u$,$d$, and $s$ quark masses.
At this order one gets the phenomenologically successful
Gell-Mann Okubo relation among meson masses. This is evidence
that the perturbative expansion in the chiral lagrangian
is good, namely that higher order terms (that modify the
Gell-Mann Okubo relation) are small.

A non--linear effective chiral lagrangian
\cite{w} is the most general
lagrangian for the baryons and the octet of quasi Goldstone
bosons (therefore valid at energies below the scale of spontaneous
chiral symmetry breaking, i.e. $E < \Lambda$) which is
compatible with the approximate accidental \suthree chiral symmetry 
of QCD \cite{g,mg}. Although there are different ways of parametrizing the
Goldstone bosons, it has been shown that they all lead to the 
same observables \cite{cwz}. This effective non-renormalizable
lagrangian consists of a power series expansion in derivatives,
baryon fields and the chiral symmetry quark mass matrix
${\cal M} = diag(m_u, m_d, m_s)$. As a trick \cite{g} to 
use chiral symmetry to also fix
the form of the explicit symmetry breaking terms using
our knowledge of QCD, the matrix ${\cal M}$ is promoted to a field
with its chiral transformation chosen so that it would fix 
the form of the ${\cal M}$ dependent terms in the QCD lagrangian
(if chiral symmetry were exact). The chiral symmetry breaking
terms in the effective lagrangian lead to s--wave interactions
of the Goldstone mesons, that are very important in the
phenomenon of kaon condensation \cite{kn,pw}. Without explicit breaking
the Goldstone bosons have only derivative couplings, which 
can be understood by recalling that the Goldstone fields
are the angular coordinates that parametrize the orbits of
degenerate vacua, so that a change in value of the field can
not affect the energy.

A chiral effective lagrangian includes all the terms compatible
with the approximate chiral symmetry of QCD with coefficients to
be determined phenomenologically when possible. Terms with  
dimension larger than four have dimensionful parameters which
are proportional to, either, inverse
powers of the symmetry breaking scale, $\Lambda \simeq
1 GeV$, or inverse powers of $f_\pi$ as shown in Eq.(\ref{nda}) below.
This prescription insures that loop corrections to each
term generate terms of the same form, if
$\Lambda \simeq 4 \pi f_\pi$ \cite{g,mg}. The usefulness
of the expansion resides in the ability of cutting the
series after a few terms. Assuming that the expansion
parameters (which are, $\partial_\mu / \Lambda$, $m_s / \Lambda$, and
$\rho_B / f_\pi^2 \Lambda = \rho_B / 7 \rho_0$, see below) are
small, the terms in the lagrangian can be organized in
successively less important sets and a finite number of 
terms can be used.
 
This method provides the only systematic way to implement the
symmetries of QCD in $\pi$--$K$--$\eta$--baryon interactions.
This method has proven to be useful, in many
applications, among which describing properties of bulk hadronic
matter, such as the formation of a kaon condensate \cite{kn,pw,bkrt}.
It has been shown
that due to the
kaon's dominant s-wave coupling to baryons \cite{pw}
the formation of a kaon condensate is quite 
insensitive to nuclear interactions.
It is believed that a kaon condensate will most likely
form at a baryonic density anywhere from 2--4 times nuclear
density.

In this paper we are not interested in the details of the onset
of the condensation but rather in solutions to the non-linear classical
field equations that describe an isolated system
consisting of baryons in a K-condensate.
In order to approach such a complicated problem
many approximations have to be made.
The result is a system of equations that resembles
 those of a liquid droplet \cite{fl,l}, 
where the baryons are treated as a gas of pseudo particles
trapped in the bose condensate of kaons \cite{lnt}.
Lynn, Nelson, and Tetradis \cite{lnt} (LNT from now on)
studied this problem using a phenomenological combination of
a chiral lagrangian, having no terms with four or more baryon fields and
no terms which are higher order than linear in quark masses,
and the Walecka lagrangian. The Walecka model \cite{cw} consists of
two fictitious massive vector and scalar fields, $V^{\mu}$ and $\phi$,
coupled to protons and neutrons in a renormalizable lagrangian.
It describes well the properties of bulk nuclear matter. LNT added the
Walecka lagrangian to incorporate nuclear forces not included
otherwise in their model. They also coupled the $\phi$ field
to the mesons. Here we only use the chiral effective
lagrangian with the four baryon terms and next to leading
order terms as well, that include more than four baryons and 
higher powers of the quark masses. In fact we have shown
elsewhere \cite{gr} that the four-fermion terms incorporate
into the chiral lagrangian the same description of bulk
nuclear matter contained in the Walecka model, when baryons
are restricted to the nucleon doublet.
Because the solutions we find have large densities,
we need not only investigate baryon-meson 
and baryon-baryon interactions but also three baryon
interactions (ie. terms with six fermion fields).
We are guided here by the belief in a perturbative series dictated
by the broken chiral symmetry as explained
above where the lowest order terms
are dominant.
 We think that this is the
essential difference between our work and LNT's treatment of
the same problem, namely we rely on a perturbative expansion as
described above and, as  we argue towards the end of this paper in
section 8, we
believe they do not.

This paper is organized as follows. The
\suthree chiral lagrangian we use is presented in section 2 and
the necessary approximations and our ansatz are given in
section 3. The following four sections discuss the
solutions. In section 4 the general method to obtain
solutions is explained using just the lowest order lagrangian.
Section 5 shows how the requirement of a continuous density
allows us to constrain the solutions to bands in  binding 
energy-density space (shown in Fig.3), with just the lowest $n = 2$
lagrangian. In section 6 and 7 we add higher order terms one
at a time to see if those terms help to obtain solutions 
with a lower density for a given binding energy. Section 7 
contains our main results. In section 8 we compare our work
with LNT and section 9 contains our conclusions.

\section*{\normalsize \bfseries{2. {The Lagrangian}}}
\addtocounter{section}{1}
  
A term in the Lagrangian consistent with n\"{a}ive 
dimensional analysis \cite{mg} is 
given by 
\begin{equation}
\label{nda}
 \rm{C} f_{\pi}^2 \Lambda^2
\left[ { \bar{B}\Gamma_{\alpha}B \over f_{\pi}^2 \Lambda} \right]^{a} 
\left[ {\overline{\Pi} \over f_{\pi}} \right]^{b}
\left[ {\partial_{\mu} \over \Lambda} \right]^{c}
\left[ {\mbox{m} \over \Lambda} \right]^{d}
\end{equation}
where $f_{\pi} = 93 \rm{MeV}$, $m = (m_{K}, m_B)$ and at first order in
$m_s$ $m_K^2 = \Lambda m_s$,
$m_s = 240MeV$, we take $m_u = m_d = 0$, and 
$\Gamma_{\alpha}$ represents 1, $\gamma_{5}$, $\gamma^{\mu}$,
$\gamma^{\mu} \gamma_{5}$, and $\sigma^{\mu \nu}$, either multiplied
by the $SU_V(3)$ generators, $T_a$, or not. The order of the
terms is given by the index $a+c+d = n$. Terms with $n = 2$ are unsupressed.
Higher orders are suppressed by $\Lambda^{2 - n}$.

The meson matrix
$\overline{\Pi}$ is defined as $\Pi_{a}T_{a}$ where $\Pi_{a}$
is the meson octet of Goldstone bosons, 
and $T_{a} \equiv \lambda_{a}/2$, where
$\lambda_{a}$ are the eight Gell-Mann matrices,
 
\begin{eqnarray}
\label{mesonmatrix}
 \overline{\Pi} = {1 \over \sqrt{2}} \left( \begin{array}{ccc}
\mbox{${\pi^{0}\over \sqrt{2}} + {\eta \over \sqrt{6}}$} & \pi^{+} & K^{+} \\
\pi^{-} & - {\pi^{0} \over \sqrt{2}} +
 {\eta \over \sqrt{6}} & K^{0} \\
K^{-} & \overline{K}^{0} & -2 {\eta \over \sqrt{6}}
\end{array} \right).
\end{eqnarray}
The baryon octet $B_{a}$ also appears in the combination
$B = B_{a}T_{a}$,

\begin{eqnarray}
\label{baryonmatrix}
 B = \left( \begin{array}{ccc}
{\Sigma^{0} \over  \sqrt{2}} + {\Lambda \over \sqrt{6}} & \Sigma^{+} & p \\
\Sigma^{-} & {- \Sigma^{0} \over \sqrt{2}} + {\Lambda \over \sqrt{6}} & n \\
\Xi^{-} & \Xi^{0} & -2 {\Lambda \over \sqrt{6}}
\end{array} \right) .
\end{eqnarray}
We consider only the octet of baryons because they are the lightest of the
baryon multiplets. 
Finally, $\rm{C}$ in Eq.(\ref{nda}) is a coefficient of O(1).
 Notice that powers
of the meson field, $\overline {\Pi}$, are not suppressed by powers
of $\Lambda$, since it appears in the combination $\overline {\Pi} / f_{\pi}$. 
The derivative factor operates on both meson and baryon fields. For
the baryon fields, however, only the spatial derivatives should be included.
We are not using a heavy baryon formalism here since the effective 
mass of the baryons will be small compared to their momentum
(see the examples provided by the values of $m^*$ in the Table 1). 
 Since we are using a Hartree approximation,
where the baryons are treated as free pseudo particles, we can use
the equation of motion to replace time derivatives by spatial
derivatives in the interaction lagrangian \cite{s}.
Notice that the density, $\rho = \bar B \gamma_0 B$,
 appears in the expansion parameter, $\rho / f_\pi^2 \Lambda
\simeq \rho / 7 \rho_0$ where $\rho_0$ is nuclear density, 
$\rho = 1.28 ~ 10^6 MeV^3$.
Therefore,
in order to keep the chiral expansion reliable we need solutions
with densities $\rho < 7\rho_0$
and we do not expect them at $\rho$ below the onset of $K$
condensation estimated at $2$--$4 \rho_0$. Thus we work in the
region in which the perturbative expansion in baryonic density is marginal.
 With $\rho = 3.4 \rho_0$ the expansion term 
$\bar{B} \Gamma_{\alpha}B / {f_{\pi}^2 \Lambda}$ is of the same order 
of magnitude as
${m_{K} / \Lambda}$.

We choose the terms in our lagrangian which satisfy the condition
$a + c + d = 2,3$. 
\begin{eqnarray}
\label{chirallag}
{\cal{L}}_{chiral} &=& {1 \over 4} f_{\pi}^2 Tr \partial_{\mu} \Sigma
\partial^{\mu} \Sigma^\dagger  
+ {1 \over 2} f_{\pi}^2 \Lambda Tr [{\cal M} \Sigma + h.c.] \nonumber\\  
\nonumber\\
&& \mbox{} + Tr \bar{B} (i \partial \sl - m_{B})B 
+ iTr \bar{B} \gamma^{\mu} [{\cal{V}}_{\mu},B] \nonumber\\
\nonumber\\
&& \mbox{} + DTr \bar{B} \gamma^{\mu} \gamma_{5} \{ {\cal{A}}_{\mu},B\}
+ FTr \bar{B} \gamma^{\mu} \gamma_{5} [{\cal{A}}_{\mu},B] \nonumber\\
\nonumber\\
&& \mbox{} - a_{1}Tr \bar{B}(\xi {\cal M} \xi + h.c.)B -
 a_{2}Tr \bar{B}B(\xi {\cal M} \xi + h.c.) \nonumber \\
\nonumber\\
&& \mbox{} - a_{3}Tr[{\cal M} \Sigma + h.c.]Tr \bar{B}B 
 + {C_{\alpha} \over f_{\pi}^2}Tr \bar{B}
 \Gamma_{\alpha}B \bar{B} \Gamma^{\alpha}B \nonumber \\ 
\nonumber\\
&& \mbox{} +
{G_{\alpha} \over f_{\pi}^4 \Lambda}Tr \bar{B}B \bar{B} \Gamma_{\alpha}B
\bar{B} \Gamma^{\alpha}B 
+ {H_{\alpha} \over f_{\pi}^2 \Lambda}
Tr [\bar{B} \Gamma_{\alpha} \partial \sl B 
\bar{B} \Gamma^{\alpha}B + h.c.] \nonumber \\
&& \mbox{} + \cdots
\end{eqnarray}
where the dots indicate terms of higher order and a few
of the same order, which involve  different
ways of contracting the baryon octet that within our 
approximations give no new terms later.
The non-linear sigma field, $\Sigma$, is given by
\begin{equation}
\label{sigma}
\Sigma = \exp{( 2i \overline{ \Pi} / f_{\pi})}  
\end{equation}
and the field, $\xi$, is defined as
\begin{equation}
\label{xi}
\xi \equiv \sqrt \Sigma = \exp{(i \overline{\Pi} / f_{\pi})}.
\end{equation}
The explicit symmetry breaking is expressed as an expansion in powers
of the small quark mass $m_{s}$, an element
of the quark mass matrix $\cal M$,

\begin{eqnarray}
\label{M}
{\cal M} = \left( \begin{array}{ccc}
0 & &  \\
& 0 & \\
& & m_{s} \end{array} \right). 
\end{eqnarray}
Finally, we have the meson vector and axial vector currents
\begin{eqnarray}
\label{vcurrent}
\cal{V}_{\mu} &=& \half (\xi^\dagger \partial_{\mu} \xi 
+ \xi \partial_{\mu} \xi^\dagger ) \nonumber\\
\label{acurrent}
\cal{A}_{\mu} &=& {i \over 2} (\xi^\dagger \partial_{\mu} \xi
 - \xi \partial_{\mu} \xi^\dagger ).
\end{eqnarray}

Most of the coefficients in Eq.(\ref{chirallag}) are
 fixed by low energy NN and $\pi N$
scattering, and by mass splittings in the baryon octet.
Their tree level values are
\begin{equation}
\begin{array}{rclrclrcl}
\label{params}
D &=& 0.81,&  F &=& 0.44,&&& \\

a_{1} &=& 0.28,&  a_{2} &=& -0.56,&  a_{3} &=& 1.3 \pm 0.2, \\

m_{s} &=& 240 \rm{MeV},& m_{B} &=& 584 \rm{MeV},&  \Lambda &\simeq& 1 \rm{GeV} 
\end{array}
\end{equation}
and, $C_{\alpha}$, $G_{\alpha}$, and $H_{\alpha}$ are free parameters of
$O(1)$.

\begin{center}
\section*{\normalsize \bfseries{3. {Ansatz and Approximations}}}
\end{center}
\addtocounter{section}{1}
\setcounter{equation}{0}

We choose a simplifying ansatz \cite{lnt} with only one non-vanishing
meson expectation value, $\langle  K^{0} \rangle$. In the presence of this
VEV the baryon masses are modified with the lightest baryon
being the neutron. For simplicity we will study the
formation of bound states with only neutrons.
We hope that this can be instructive in searching 
for solutions with more degrees of freedom.
Our ansatz for the classical meson expectation value is, therefore,

\begin{equation}
\label{ansatz}
 \overline{\Pi} = {1 \over \sqrt{2}} \left( \begin{array}{ccc}
0 & 0 & 0  \\
0 & 0 & \langle  K^{0} \rangle \\
0 & \langle  \overline{K}^{0}\rangle & 0 \end{array} \right) 
\end{equation}
where $\langle K^{0}\rangle$ is chosen to be real, and from 
the hermiticity of
$\overline{\Pi}$ we get
 $\protect{\langle  \overline{K}^{0}\rangle = \langle K^{0}\rangle}$. 

We are interested in finding classical
solutions of the equations of motion consisting of bound
states at zero pressure and with large baryon number. 
In order to minimize their surface energy these solution
should be spherically symmetric. Furthermore, we will consider
bound states large enough for the surface effects to be
negligible. We take 
the density of baryons and the value of the meson
field to be almost constant throughout the interior.
Using the Thomas-Fermi approximation \cite{b}
we assume that the kaon fields vary slowly compared to 
the baryon wavelength, i.e. $\vec{\nabla}K^0/f_{\pi} \ll \vec{k} $, where
$\vec k$ is the neutron's wavenumber. 
This allows us to effectively treat
the neutrons at each point as a Fermi gas in a constant
kaon field. Finally, since the kaon field is slowly
varying we find that all p-wave and higher derivative interactions
are negligible compared to s-wave interactions. With the ansatz given
in Eq.(\ref{ansatz}), the vector current, $\mathcal{V}_{\mu}$, vanishes and
the terms with axial vector coupling---$D$ and $F$ terms---to the 
neutrons given in Eq.(\ref{chirallag})
also vanish.
With these simplifications the lagrangian reduces to
 \begin{eqnarray}
\label{rcl}
{\cal{L}}_{chiral} &=& \half f_{\pi}^2 \partial_{\mu} \theta
\partial^{\mu} \theta - f_{\pi}^2 m_{K}^2 (1- \cos{\theta}) \nonumber\\
\nonumber\\
&& \, + \bar{n} (i \partial \sl - m_{n}^{*})n
- {C_{V}^2 \over 2f_{\pi}^2} n^{\dagger}nn^{\dagger}n \nonumber\\
\nonumber\\
&& \,+ {C_{S}^2 \over 2f_{\pi}^2} \bar{n}n \bar{n}n
+ {G_{1} \over 3f_{\pi}^4 \Lambda} \bar{n}n \bar{n}n \bar{n}n \nonumber\\
\nonumber\\
&& \, + {G_{2} \over f_{\pi}^4 \Lambda}n^{\dagger}nn^{\dagger}n \bar{n}n
+ {H_1 \over f_{\pi}^2 \Lambda} \bar{n} i \vec{\gamma} \cdot \vec{\nabla}
n \bar{n}n
\end{eqnarray}
where $\theta = {\sqrt{2} \langle K^{0} \rangle / f_{\pi}}$,
we have used the first order (in $m_s$) relation 
$m_K^2 = m_s \Lambda$,
$n$ is the neutron field,  $m_n^*$ is the
neutron mass within the $\langle K^0 \rangle$ condensate \cite{lnt},
\begin{equation}
\label{mns} 
m_{n}^{*} = m_{N} - a_t m_s(1 - \cos{\theta}),
\end{equation}
with $a_t = 2 a_3 +a_2 + a_1 = 2.32 \pm 0.4$
and $m_N$ the free neutron mass, $m_N = 939 MeV$. Notice that
a constant has been added to the lagrangian in Eq.(\ref{rcl})
so that the $\theta$ dependence of ${\cal{L}}_{chiral}$ disappears
from the lagrangian when $\theta = 0$. This is the
origin of the $(1 - \cos {\theta})$ factor in the second
term of ${\cal{L}}_{chiral}$ in Eq.(\ref{rcl}). Notice that there
is also
a factor $(1 - \cos {\theta})$ in $m_n^*$, Eq.(\ref{mns}), so 
that $m_N$ is the physical neutron mass (outside the condensate). 
The constant $m_N$ contains, therefore, all contributions of
the form $Tr \bar{B} B$ in Eq.(\ref{chirallag}) obtained by
setting $\xi = 1$ and $\Sigma = 1$, thus
$m_N = m_B + 2 m_s(a_3 + a_2)$ (actually, knowing $a_2$ and
$a_3$ this relation fixes $m_B$, Eq.(\ref{params})).

 The three vector momentum and spin dependent terms containing
currents of the form $\bar n \gamma^i n$ and $\bar n \gamma^i \gamma_5 n$,
 where
$i$ are the spatial degrees of freedom, average to zero in 
spherically symmetric bulk matter. Furthermore, all terms
containing time derivatives of the condensate vanish as
a result of minimizing the thermodynamical potential at 
fixed electric charge: the time dependence turns out to be simple
harmonic with the frequency equal to the electric charge \cite{kn,pw}.

\section*{\centering \normalsize \bfseries{4. {Infinite Solutions with Lowest
Order $\mathbf{n = 2}$ Terms}}}
\addtocounter{section}{1}
\setcounter{equation}{0}

Let us first analyze the $n = 2$ terms of Eq.(\ref{rcl})
($G_1 = G_2 = H_1 = 0$). \linebreak
The Thomas-Fermi approximation allows one to take a free particle
wave function,
 $\! \exp {i(\vec{k} \cdot \vec{x} - \epsilon_k t)}$, for the neutron,
where $\vec k$ and
$\epsilon_k$ are the space dependent wavenumber and energy respectively.
The Dirac  
equation for a neutron moving in the mean field of all
other neutrons is    
\begin{eqnarray}
\label{dirac1}
\left[ 
i  \partial \sl - m_{n}^{*} - {C_{V}^2 \over f_{\pi}^2} (n^{\dagger}n) \gamma^{0}
+ {C_S^2 \over f_{\pi}^2} (\bar{n}n) \right] n = 0,
\end{eqnarray} 
and by substituting in  the neutron wave function, we get
\begin{eqnarray}
\label{dirac2}
\left[
\gamma^0 ( \epsilon_k - {C_{V}^2 \over f_{\pi}^2} (n^{\dagger}n))
- \gamma^i k_i - (m_{n}^{*} - {C_S^2 \over f_{\pi}^2} (\bar{n}n))
\right] n =0.
\end{eqnarray}
Squaring the above equation and applying Dirac algebra we
get the dispersion relation of a free quasi-particle,
\begin{equation}
\label{disrel}
\epsilon^{*2} - k^2 - m^{*2} = 0,
\end{equation}
where we define the effective energy and mass of the quasi-particle to be
\begin{eqnarray}
\label{effe}
\epsilon^{*} &\equiv& \epsilon_{k} - {C_V^2 \over f_{\pi}^2}n^{\dagger}n \\
\label{effm}
m^{*} &\equiv& m_{n}^{*} - {C_S^2 \over f_{\pi}^2} \bar{n}n.
\end{eqnarray}

The zero temperature ground state, $| \Psi \rangle$,
 of our system can be
found by minimizing the thermodynamic potential, $\Omega$. For fixed
baryon number, $B$,
\begin{equation}
\label{thermo}
\Omega = E_{total} - \mu_{B} B,
\end{equation}
where $\mu_B$ is the baryon chemical potential and 
 $B = \int d^{3}x n^{\dagger}n$. The total energy of the system, $E_{total}$,
is given by
\begin{equation}
\label{etot}
 E_{total} = T^{00} = \int d^{3}x (\varepsilon_k + \varepsilon_n +
{\cal U}_\theta)
\end{equation}
 where $\varepsilon_k$ is the kinetic
energy density of the condensate,
\begin{equation}
\label{kencon}
\varepsilon_k = \half {f_\pi}^2 ({\dot{\theta}}^2 + 
(\vec{\nabla} \theta)^2) \nonumber \\
= \half f_{\pi}^2(\vec{\nabla} \theta)^2,
\end{equation}
$\varepsilon_n$ is the neutron energy density,
\begin{equation}
\label{eden}
\varepsilon_n =  2 \int_{}^{k_f}{{d^{3}k \over
 (2 \pi)^{3}}(k^2 + m^{*2})^{\half}}
+ {C_{V}^2 \over 2f_{\pi}^2}(n^{\dagger}n)^2 
 +{C_{S}^2 \over 2f_{\pi}^2}( \bar{n}n)^2,
\end{equation}
and ${\cal U}_\theta$ is the potential energy density
of the condensate
\begin{equation}
\label{U}
{\cal U}_\theta = f_{\pi}^2 m_{K}^2(1 - \cos{\theta}).
\end{equation}

The thermodynamic potential, $\Omega$, is a function of 
the Fermi momentum, $k_{f}$, the $\langle K^0 \rangle$
condensate, $\theta$, and
$\mu_{B}$ as well as the other coefficients in the Lagrangian. First we
functionally minimize $\Omega$ with respect to $k_f(\vec{x})$, and find
\begin{equation}
\label{mub}
\mu_{B} = \mu^* + {C_V^2 \over f_{\pi}^2}n^{\dagger}n,
\end{equation}
where $\mu^*$ is the quasi-particle's chemical potential
defined as the value of $\epsilon^*$ at $k = k_f$. Eq.(\ref{mub}) is 
equivalent to Eq.(\ref{effe}) with $k = k_f$, hence, $\mu_{B}$ is equivalent to
the energy of a neutron at the top of the Fermi sea, which is
therefore
constant over all space.
 Different choices for the chemical
potential lead to different sizes
of finite solutions, i.e. to different numbers of neutrons within it.

Functionally minimizing $\Omega$ with respect to $\theta(\vec{x})$ 
results in a differential
equation for the condensate, $\theta$,
\begin{equation}
\label{soliton}
\vec{\nabla}^2(f_{\pi} \theta) - {\partial \over \partial(f_{\pi} \theta)}
\left[ P_{n} - {\cal U}_\theta  \right] = 0,
\end{equation}
where $P_{n}$, the pressure of the neutron gas, 
is given by  

\begin{eqnarray}
\label{pressure1}
P_{n} &=& \mu_B n^{\dagger}n - \varepsilon_n \\
\label{pressure2}
&=& {1 \over 3}T^{ii} =
(\bar{n} \vec{\gamma} \cdot \vec{k} n)
+ {C_V^2 \over 2 f_{\pi}^2}(n^{\dagger}n)^2
- {C_S^2 \over 2 f_{\pi}^2}(\bar{n} n)^2.
\end{eqnarray}

In the zero temperature ground state the neutron number density is
\begin{equation}
\label{density}
\rho_{n} = \langle \Psi | n^{\dagger}n | \Psi \rangle = 
2 \int^{k_f}{d^3k \over (2 \pi)^3} = {k_f^3 \over 3 \pi^2},
\end{equation}
the scalar density
\begin{eqnarray}
\label{scalarden}
\rho_{S} &=& \langle \Psi | \bar{n}n | \Psi \rangle =
2\int^{k_f}{d^3k \over (2 \pi)^3}{m^* \over \sqrt{k^2 + m^{*2}}} \nonumber\\
\nonumber\\
&=& {m^* \over 2 \pi^2} \left[ k_f \mu_B - {{m^*}^2 \over 2} \ln {
\left( {\mu_B + k_f \over \mu_B - k_f} \right)} \right],
\end{eqnarray}
and
\begin{eqnarray}
\label{vectorden}
\langle \Psi |(\bar{n} \vec{\gamma} \cdot \vec{k} n)| \Psi \rangle &=&
2 \int^{k_f}{{d^3k \over (2 \pi)^3}{k^2 \over \sqrt{k^2 + m^{*2}}}} \nonumber\\
\nonumber\\
&=& {3 \over 4 \pi^2} \left[ \mu_B \left( {k_f^3 \over 3} - {k_f {m^*}^2 \over 2}
\right) +
 {{m^*}^4 \over 4} \ln {\left( {\mu_B + k_f \over \mu_B - k_f} \right)} \right].
\end{eqnarray}

Substituting Eqs.(\ref{effm}), (\ref{mub}), (\ref{density}),
and (\ref{scalarden}) into the equation for the dispersion
relation at the top of the Fermi sea,
\begin{equation}
\label{disper}
{\mu^*}^2 - k_f^2 = {m^*}^2
\end{equation}
we get a  a transcendental
 equation of the variables
$k_f$ and $\theta$ which we solve numerically to get
$k_f(\theta)$.
We can now solve Eq.(\ref{soliton}) after
plugging $k_f(\theta)$ into Eq.(\ref{pressure2})
to obtain $P_n(\theta)$ .

We can view Eq.(\ref{soliton}) as a one dimensional newtonian 
equation of motion \cite{l} 
for a particle of unit mass moving in a
potential, $V_{\mathit{eff}}$, with the
following replacements:

\begin{equation}
\label{replacement}
f_{\pi} \theta \rightarrow x, \hspace{1cm}
r \rightarrow t, \hspace{1cm}
\left[ P_n - {\cal U}_\theta \right] \rightarrow V_{\mathit{eff}}(x).
\end{equation}
Writing Eq.(\ref{soliton}) in radial coordinates, we get (for our spherically
symmetrical, time independent condensate $\theta$)

\begin{equation}
\label{diff}
{d^2(f_{\pi} \theta) \over dr^2} + {2d(f_{\pi} \theta) \over rdr} =
-{d(P_{n} - {\cal U}_\theta) \over d(f_{\pi} \theta)}.
\end{equation}

Notice that the ``damping term'' decreases as $1/r$ which means
it becomes negligible for large ``times'' in the newtonian analogy.
The potential, $V_{\mathit {eff}} = P_n - {\cal U}_\theta$ for an
infinite baryon solution, with
$C_V^2 = .24$, $C_S^2 = .63$, and $\mu_B = 900 \rm{MeV}$, is shown in Fig.1.
The potential has two degenerate maxima one at $\theta_0 = 1.37$ and the
other at the true vacuum, $\theta = 0$. In the newtonian analogy 
a test particle starts from the top of the hill at $\theta_0 = 1.37$ and
waits there for a very long time during which the
damping term becomes effectively zero. The particle then
accelerates quickly through the
valley reaching the top of the other hill, the true vacuum,
where it comes to a stop. Therefore, the  solutions for a
large number of baryons have constant $\theta$ and density
 over a large volume and a small spherical surface region
 where one vacuum state evolves 
rapidly to the other. One can isolate the infinite  solutions
by imposing the following conditions on $V_{\mathit {eff}}$,
\begin{eqnarray}
\label{infinite}
V_{\mathit {eff}}(\theta_0) &=& 
V_{\mathit {eff}}(\theta = 0) \, = \, 0 \nonumber\\
\nonumber\\
\left. \frac{dV_{\mathit {eff}}(\theta)}{d\theta} \right |_{\theta_0} &=& 0.
\end{eqnarray}
where $\theta_0$ is the value of the condensate at the center
of the solution, $r = 0$.
Notice that for an infinite solution $V_{\mathit{eff}}(\theta_{0}) =
\mu_B n^{\dagger}n - \varepsilon_n - {\cal U}_{{\theta}_{0}} = 0$ and
in the constant interior of the solution $\varepsilon =
\varepsilon_n + {\cal U}_{{\theta}_0}$ giving the relation,
\begin{equation}
\label{bind}
\mu_B = {\varepsilon \over \rho_n},
\end{equation}
thus the energy per neutron is $\mu_B$.

The strangeness density of each solution is found as the
zero component of the strangeness current, $S_{\mu} =
 (J_Y)_{\mu} - (J_B)_{\mu}$, where $J_Y$ is the strong
hypercharge current and $J_B$ is the baryon number
current. With our approximations the only non zero contribution
to $S_0$ comes from the meson vector current coupling
$ iTr \bar{B} \gamma^{\mu} [{\cal{V}}_{\mu},B]$ term
in Eq.(\ref{chirallag}), and it is $S_0 = (1 - \cos \theta_0)
n^{\dagger} n$ so the strangeness per baryon number
of the strange matter is just $(1 - \cos \theta_0)$ (see Table 1).

\begin{center}
\section*{\normalsize \bfseries{5. {Mapping Out the Solution Space}}}
\end{center}
\addtocounter{section}{1}
\setcounter{equation}{0}

The model we are considering so far with only
$n = 2$ terms, which is equivalent to the 
Walecka model 
for bulk nucleonic matter \cite{cw},
 has only two parameters, $C_V^2$ and $C_S^2$. Notice that the
coefficients $C_S^2$ and $C_V^2$ are positive definite and the 
signs of the terms are chosen to account for the scalar
attraction and vector repulsion observed in nuclear interactions.
Thus we have four unconstrained parameters, $C_S^2$, $C_V^2$, $\mu_B$,
and $\theta_0$, and 
two constraining conditions on $V_{\mathit {eff}}$, Eq.(\ref{infinite}).
We are left with two independent variables that we choose
to be $\mu_B$ and $C_S^2$.
Since the main properties
of the infinite solutions we are looking for are their baryon
density, $\rho_n$, and their binding energy, $E_b$, we will
show the regions of solutions in the $(\rho_n, E_b)$ space. This
is equivalent in the variables we are talking about to a
$(\theta_0, \mu_B)$ space. This is so because $\rho_n =
k_f^3 / 3 {\pi}^2$ depends on $\theta_0$ through the
monotonically increasing (see below) function, $k_f(\theta_0)$,
and $E_b = \mu_B - m_N$, see Eq.(\ref{bind}), inside the
infinite solution, i.e. in bulk baryonic matter. For a given
$\mu_B$, i.e. a given $E_b$, there is a range of allowed
number densities, that depend on $C_V^2$ and $C_S^2$, $\rho_n =
\rho_n(C_V^2, C_S^2)$, but $C_V^2$ also depends on $C_S^2$, 
$C_V^2 = C_V^2(C_S^2)$. $C_V^2$ can be shown numerically to
\emph{increase} monotonically with $C_S^2$, while $\rho_n$
\emph{decreases} monotonically with $C_S^2$. So, the
allowed range of $\rho_n$ for each $E_b$ corresponds to an
allowed range of $C_S^2$ values, $0 \leq C_S^2 \leq (C_S^2)_{max}$.
It turns out $(C_S^2)_{max}$ can be found in a systematic way,
that we pass now to explain.

Let us return to Eq.(\ref{disper}), the transcendental  
equation whose roots we solve numerically for to find
$k_f$ at a fixed $\theta$. Using Eqs.(\ref{mub}) and
(\ref{density}) we see that $\mu^* = \mu_B -
({C_V}^2 / f_{\pi}^2)(k_f^3 / 3 {\pi}^2)$, therefore
the l.h.s. of Eq.(\ref{disper}) does not explicitly depend on
$\theta$. Through Eqs.(\ref{effm}) and (\ref{scalarden}) we
see that the r.h.s., ${m^*}^2$, carries the explicit $\theta$
dependence of Eq.(\ref{disper}) through $m_n^*$. The l.h.s.
depends on $\mu_B$, $C_V^2$, and $k_f$ and the r.h.s. depends
on $C_S^2$ and $k_f$. Figs.2a and 2c are examples of the sides 
of Eq.(\ref{disper}) as functions of $k_f$ for particular
$C_S^2$ and $C_V^2$ values. The l.h.s are shown with black 
dashed lines for a given $\mu_B$ ($\mu_B = 900MeV)$, the
r.h.s. are shown for different $\theta$ values with gray solid lines. The
shape of the gray solid curves depends in a complicated way on
$C_S^2$. For each $\theta$, the intersections of the corresponding gray line
with the dashed line gives the solution $k_f(\theta)$. The
dashed line moves up and down the diagram with $\mu_B$ without
changing shape. For a fixed $\mu_B$, as $\theta$ increases
from zero the solution $k_f(\theta)$ starts departing
from zero and grows continuously to a maximum value where
all the gray lines turn over, thus $k_f(\theta)$ is a 
monotonically increasing function. This happens when there 
is only one intersection (for $\mu_B$ and $\theta$ fixed).
This is the case of Fig.2a. However there are cases, such as the one
shown in Fig.2c, in which there are three intersections
(see the three x's). We can see in Fig.2c that as $\theta$
increases (gray line lowers with respect to the dashed line) the
first two intersections approach each other, get to coincide at the point 
where the two intersecting curves have the same slope, and
disappear, leaving only the third intersection. If we had 
taken the $1^{st}$ intersection as the physical $k_f(\theta)$,
at the $\theta$ value for which the two first intersections
disappear, after joining in one point, there would be a
discreet jump in $k_f$ from this point to the $3^{rd}$ intersection,
at a larger value of $k_f$, which becomes now the only intersection.
This jump in $k_f$ is unphysical, since the density $\rho_n \sim
k_f^3$ has to be a continuous function of the condensate,
$\theta$, as $\theta$ grows from zero outside the bound state to
its maximum value in the interior. Thus we impose conditions
on the l.h.s. and r.h.s. of Eq.(\ref{disper}) as
functions of $k_f$ to forbid the values of $C_S^2$ (for each 
$\mu_B$) for which triple intersections appear.
These are conditions on the slope of the intersecting curves
(i.e. the l.h.s and the r.h.s.) as functions of $k_f$.
Valid solutions only occur when at $k_f = 0$ the r.h.s,
${m^*}^2$ (gray line) is lower than the l.h.s., $( {\mu^*}^2 - k_f^2)$
(black dashed line), see Figs.2a and 2c. This insures that 
as the condensate $\theta$ increases (gray line lowers)
$k_f(\theta)$ (and the baryon density) increases smoothly
from zero near the surface of the solution\footnote{Notice
that if instead  the black curve is  lower than the gray
curve at $k_f = 0$, as $\theta$
increases the first intersection happens at $k_f \neq 0$
and approaches $k_f = 0$. This corresponds to the unphysical
situation of a discontinuous density that starts all of a
sudden with a large value near the surface and decreases
at the center of the solution.}.
The curves cross when they
intersect and ${m^*}^2$ becomes larger. If the ${m^*}^2$
curve is more concave than the $({\mu^*}^2 - k_f^2)$ one,
they cross again. This is the case we reject, namely
when
\begin{equation}
\label{dercond}
{d \over d k_f} \left( {m^*}^2 - {\mu^*}^2 + k_f^2 \right)
< 0.
\end{equation}
for any value of $\theta$ and $k_f$.
Notice that both slopes are negative and we reject the case in which the 
${m^*}^2$ slope is larger (more negative) than the other.
We show the difference in slope
for a rejected case (the one of Fig.2c) in Fig.2d.
Fig.2b shows an allowed border line case (the one of Fig.2a) in
which the difference in slope would become negative (for some
values of $k_f$ and $\theta$) if $C_S^2$ would be increased.
This is the important numerical result that allows us to get
an upper bound on $C_S^2$, the fact that the difference
in slopes (i.e. the l.h.s. of Eq.(\ref{dercond})) decreases
with increasing $C_S^2$. Thus, the border line case
for each $\mu_B$ fixes $(C_S^2)_{max}$.

Now we can explain the procedure actually
followed to map out the solution space. First, we
choose a $\mu_B$ (or equivalently $E_b$). Then vary
$C_V^2$ and $C_S^2$ and look at $V_{\mathit{eff}}$ 
(solving Eq.(\ref{disper}) to get $k_f(\theta))$ until
the conditions for an infinite solution, Eq.(\ref{infinite}),
are fulfilled, namely until the maximum of $V_{\mathit{eff}}$
is at zero. This is not difficult to do
after noticing that increasing $C_S^2$ (scalar attraction)
raises $V_{\mathit{eff}}$ and increasing $C_V^2$ (vector repulsion)
lowers $V_{\mathit{eff}}$. Once a set $(C_S^2, C_V^2)$
corresponding to an infinite solution has been found,
we look at the l.h.s. of Eq.(\ref{dercond}) (i.e. the difference
of the slopes of both sides of Eq.(\ref{disper})). If
it corresponds to an allowed case (positive difference in slope)
$C_S^2$ is increased, otherwise $C_S^2$ is decreased,
and the shape of $V_{\mathit{eff}}$ has to be checked again
to obtain another $C_V^2$ so that the new $(C_S^2, C_V^2)$
corresponds to an infinite solution. One keeps doing this 
until finding an infinite solution corresponding to a
border line case (Fig.2b) for the difference in slope.
At this point we have found $(C_S^2)_{max}$ for the given
$\mu_B$, and its corresponding $C_V^2$ and $\theta_0$. This procedure
determines the $({\rho_n})_{min}$ border of the allowed regions
in $( E_b, \rho_n)$ space. The $({\rho}_n)_{max}$ border
corresponds to $C_S^2 = 0$ and the $C_V^2$ value necessary
to satisfy Eq.(\ref{infinite}), for each $\mu_B$. The allowed 
regions found are shown in Fig.3 with labels $a_3 = 1.3$
and $a_3 = 1.5$. The parameter $a_3$ appears in ${m_n}^*$, 
Eq.(\ref{mns}). It is measured through the nucleon
$\sigma$-term to be $a_3 = 1.3 \pm 0.2$. It is
the parameter responsible for the s-wave attraction between
kaons and nucleons, that yield K-condensation. We see in Fig.3 that
a higher value of $a_3$, i.e. a lower value of the
effective nucleon mass, $m_n^*$, helps to find 
solutions at lower densities. Only values of $C_V^2 \geq 0$ are
allowed, producing the lower boundaries shown in the figure with the
$C_V^2 = 0$ label. Some values of $(C_V^2, C_S^2)$ are also shown.

Notice that the procedure we follow is only self
consistent for $\rho_n / \rho_0 < 7$, since the perturbative
expansion in baryon bilinears $\bar{B} \Gamma_{\alpha} B$ is
lost at larger densities. The rest of the diagram for $\rho_n
> 7 \rho_0$ in Fig.3 may be indicative but is not believable.
In any case, we were not expecting solutions below the density
necessary for the onset of K-condensation, $2$--$4$ $\rho_0$.
This is consistent with the densities we find, starting at 
$4.5 \rho_0$. At these densities, the fermion bilinear expansion
parameter $\sim \rho_n / 7 \rho_0$ is large and higher order
terms with more baryons should be considered. 

\section*{\centering\normalsize \bfseries{6. {Solutions with Higher Order
$\mathbf{n = 3}$ Terms}}}
\addtocounter{section}{1}
\setcounter{equation}{0}

Let us describe the effect of taking
into account the $G_1$, $G_2$, and $H_1$ terms in
${\cal L}_{chiral}$, Eq.(\ref{rcl}).
These are the only remaining terms of order $n = 3 \, (a + c = 3)$,
besides the $a_1, a_2, a_3$ terms $(a + d = 3)$, already
included in our initial lagrangian, because they are the
lowest order terms providing an s-wave (i.e. derivative or
momentum independent) K-B couplings, thus without
them kaons and baryons would not be coupled within bulk
baryonic matter. 

The procedure we follow to examine the effect of these higher order terms
is to add only one term at a time. This increases by one the
number of free parameters, $\mu_B$, $\theta_0$, $C_V^2$, $C_S^2$.
However we reduce the problem again to just the old four parameters 
by fixing the new one. We choose both negative and positive
values of $O(1)$ for the single higher order 
parameter studied, within a range that insures the
higher order term in the lagrangian is never larger than
the $C_V^2$ and $C_S^2$ terms. Then we proceed
 as before choosing a $\mu_B$ value and finding $(C_S^2)_{max}$
(and the corresponding $(\rho_n)_{min}$, etc.) as described
above.

We are interested in knowing which higher order terms help to
get solutions with lower density $\rho_n$ for a given binding
energy. The three baryons $G_1$ and $G_2$ terms 
raise $(\rho_n)_{min}$ instead,
as well as the $H_1$ term with positive $H_1$ values,
 while with negative $H_1$ values it
can lower $(\rho_n)_{min}$ although very little.
An example of the effect
of these terms is shown for $G_1 = 0.4$ (and $a_3 = 1.3$) in
Fig.3. The region of allowed solutions in $(\rho_n, E_b)$ is
a wedge because $(C_S^2)_{max}$ becomes zero at the
tip (and the $(\rho_n)_{max}$ boundary corresponds to $C_S^2 = 0$).
Other examples with $G_1$, $G_2$, and $H_1$ positive and 
negative are given in Table 1. 
This table shows the values of $(C_S^2)$, $(C_V^2)$, the one higher order
parameter chosen to be non-zero, the baryon number density,  the
effective mass $m^*$ of the quasi particles and  the
condensate $\theta_0$, for which the baryon number density is minimum
(corresponding to the  maximum values of $(C_S^2)$ and $(C_V^2)$ for 
which solutions exist, after fixing the higher order parameter to the
largest positive and negative physically acceptable values)
for a fixed binding energy $E_b=-39MeV$ and parameter $a_3=1.3$.
 While $G_1$ does not introduce
any major change in the self consistent procedure 
described above used to find
solutions, the $G_2$ and $H_1$ terms turn the
$\mu^*$ algebraic Eq.(\ref{mub}) into 
a transcendental equation, (see Eq.(A.1) in the appendix with either
$H_1 = I_V = 0$ or $G_2 = I_V = 0$). The appendix
gives the complete equations with all the higher order
terms included in this paper for the effective neutron
chemical potential, effective neutron mass, pressure
and energy density that we label with the sub index
``$_{TOTAL}$''. Notice that the $H_1$ term in Eq.(\ref{rcl})
contains a derivative of the nucleon field so it 
modifies the momentum of the neutrons. Adding only $H_1$
to the $C_V^2$ and $C_S^2$ parameters, Dirac's equation
becomes
\begin{eqnarray}
\label{dirac3}
\left[
\gamma^0 \epsilon^* - (1+{H_1 \over f_\pi^2
 \Lambda}(\bar n n))
\vec{\gamma} \cdot \vec k  
 - (m^{*} - {H_1 \over f_{\pi}^2 \Lambda} (\bar{n}\vec{\gamma} \cdot 
\vec{k} n)) 
\right] n =0.
\end{eqnarray}
where $\epsilon^*$ and $m^*$ are those found in Eqs.(\ref{effe})
and (\ref{effm}). One can arrive at a standard free particle
dispersion relation by dividing Eq.(\ref{dirac3}) by the
factor multiplying $\vec{\gamma} \cdot \vec{k}$ and redefining the
effective energy and mass. This is the origin of the
$(1 + (H_1 / f_{\pi}^2 \Lambda) \bar{n} n)$ denominator
in $\mu^*_{\scriptscriptstyle{TOTAL}}$ 
(i.e. ${\epsilon^*}_{\scriptscriptstyle{TOTAL}}$ at the top
of the Fermi sea, where $k = k_f$) 
and $m^*_{\scriptscriptstyle{TOTAL}}$ given
in the appendix Eqs.(A.1) and (A.2). 

The importance of having to solve 
a transcendental equation for $\mu^*(k_f)$ when $G_2$ or
$H_1$ are non-zero is that many of their solutions are 
incompatible with the existence of infinite solutions. The reason
is that solutions for $\mu^*(k_f)$ can not be found for large
ranges of $G_2$ or $H_1$ for ${\mu^*}^2(k_f) < k_f^2 - \Delta$, 
where $\Delta$ depends on $G_2$ or $H_1$ and $C_V^2$.
Since ${\mu^*}^2 - k_f^2 = {m^*}^2$ (where $\epsilon^*$ and
$\mu^*$ are 
$\epsilon^*_{\scriptscriptstyle{TOTAL}}$ 
and $m^*_{\scriptscriptstyle{TOTAL}}$ given in the appendix with
only $C_S^2$, $C_V^2$, and $G_2$ or $H_1$ respectively non-zero)
this means that the effective mass has a non-zero minimum
${m^*}^2 > \Delta = m^{*2}_{min} > 0$. This forbids
solutions that would require ${m^*}^2 < \Delta$ in the bulk matter.
Since $\Delta$ increases with increasing values (positive and negative)
of $G_2$ and $H_1$, only small values of these parameters lead to 
solutions, and even then $(\rho_n)_{min}$ is not improved.

From this analysis we conclude that pure nuclear interaction terms
do not help to lower $(\rho_n)_{min}$ for a given binding energy.
Then, one is lead to try higher order terms dependent on the
condensate. As we will see they may actually help.

\begin{center}
\section*{\normalsize \bfseries{7. {Solutions with Higher Order
Condensate Dependent Terms}}}
\end{center}
\addtocounter{section}{1}
\setcounter{equation}{0}

The first terms of this type are of order $n = a + d = 4$.
They are obtained by squaring the trace in the $2^{nd}$ term
of ${\cal{L}}_{eff}$ in Eq.(\ref{chirallag}), by multiplying
the $a_1$, $a_2$, and $a_3$ terms in Eq.(\ref{chirallag}) by
$\bar{B} B / f_{\pi}^2 \Lambda$, by multiplying the
$C_V^2$ and $C_S^2$ terms in Eq.(\ref{chirallag}) by
$Tr [ {\cal{M}} \Sigma + h.c.]$ and by writing similar
terms with alternative ways of taking the traces. They all give
only the following three terms within bulk matter with only neutrons in a
$K^0$ condensate,
\begin{eqnarray}
{\cal{L}}_4 & = & {I_S \over 2 f_{\pi}^2 \Lambda} m_s (1 - \cos{\theta})
\bar{n}n\bar{n}n - {I_V \over 2 f_{\pi}^2 \Lambda} m_s (1 - \cos{\theta})
n^{\dagger}nn^{\dagger}n \nonumber\\
\nonumber\\
&& \,- J f_{\pi}^2 m_s^2
(1 - \cos^2{\theta}).
\end{eqnarray}
The terms $I_S$ and $I_V$ effectively add a $\theta$ dependence
to the ${C_S}^2$ and ${C_V}^2$ terms, respectively.
While $I_V$, that modifies $\mu^*$ (see Eq.(A.1)), does
not help to lower $\rho_n$, positive values of $I_S$, that
modify $m^*$ (see Eq.(A.2)), do help (see Table 1 and
notice that $(\rho_n)_{min} = 5.3 \rho_0$ for $I_S = 3$
instead of $6.5 \rho_0$ for $I_S = 0$).
The $J$ term helps in lowering $\rho_n$, and it is
experimentally determined, because it corrects the 
kaon mass and other terms in the potential energy, $\cal{U}_{\theta}$
(see Eq.(A.5)). We must be careful since now 
$m_K^2 = m_s \Lambda + ...$ because we include higher order terms in $m_s$.
Let us now call $\tilde{m}_{K}$ the value of the K-mass at first order
in $m_s$, namely $\tilde{m}_K^2 = m_s \Lambda$. 
Therefore we change the notation which we used in Eq.(\ref{rcl}) for
the term $f_{\pi}^2 m_K^2 (1- \cos{\theta})$
into $f_{\pi}^2 \tilde{m}_K^2 (1- \cos{\theta})$, and
we write ${\cal{U}}_{\theta}$
as
\begin{eqnarray}
\label{pot}
{\cal U}_{\theta} &=& f_{\pi}^2 \tilde{m}_{K}^2 (1 - \cos{\theta}) +
J {f_{\pi}^2 \over \Lambda^2} \tilde{m}_{K}^4 (1 - \cos^2 \theta) + \cdots.
\end{eqnarray}
In order to evaluate $\tilde{m}_K$, the $1^{st}$
order value of the kaon mass,  we use the Gell-Mann-Okubo
relation, that holds only at first order in the quark masses,
 $\tilde{m}_{K}^2 = 1/4(3 m_{\eta}^2 + m_{\pi}^2)$,
 where we take $m_{\pi}$ and
$m_{\eta}$ to be the physical masses of the $\pi$ and $\eta$ mesons
respectively.  We find $\tilde{m}_{K}^2 = m_s \Lambda \simeq
(480 MeV)^2$. We take this as a rough estimate of the minimum 
value of 
$\tilde{m}_K$.\footnote{Taking $\tilde{m}_K$ instead of the
 physical $m_K$ mass
to be equal to $m_s \Lambda$ amounts to  a redefinition of the strange quark
mass $m_s$. This, however, would not effect our previous results
because they do not depend explicitly on $m_s$, but
rather on such combinations as $m_s a_3$, $m_s I_S$, etc, and the change
in $m_s$ can be compensated by small changes in the accompanying parameters.}

 Expanding $\cos^2 \theta$ and leaving only the
terms quadratic in $\theta$, $\theta^2 = 2 \langle K^0 \rangle^2
/ f_{\pi}^2$, Eq.(\ref{pot}) becomes
\begin{eqnarray}
\label{uchiral}
{\cal U}_{\theta} &=&  m_K^2 {\langle K^{0}\rangle}^2 \left(
\left({{\tilde m}_K \over m_K} \right)^2 +
 {2J {{\tilde m}_K}^4
\over {\Lambda}^2 m_K^2} + \cdots \right) + \cdots \nonumber\\
 &=& m_K^2 {\langle K^{0}\rangle}^2 
(0.94 + .43 J + \cdots) + \cdots,
\end{eqnarray} 
where  $m_K$ is the physical kaon mass, $m_K = 498MeV$.
Since the parenthesis must be 1, we get an upper bound
on $J$, $J < 0.14$.

In order to examine the most favorable case, the lowest $\rho_n$ for
a given $E_b$, we take, besides the largest reasonable value for $a_3 $
i.e. $a_3 = 1.5$,  $J = 0.14$ and $I_S$ as large as possible,
 while keeping the
$I_S$ term smaller than the $C_S^2$ term in agreement with a 
perturbative expansion. We show the results in Fig.3.
Because the largest value of $\theta_0$ encountered in the
corresponding solutions is $\theta_0 \simeq 1.5$, the 
value $I_S = 4.6 C_S^2$ is the largest insuring the $I_S$ term
is never larger than the $C_S^2$ term. These solutions, those
with the lowest $\rho_n$ we find for every
binding energy, are our main result. We believe the perturbative
chiral lagrangian cannot reasonably do any better.
Notice that solutions with $\rho_n < 4 \rho_0$ have at most
only $50MeV$ of binding energy, and those with $\rho_n < 5 \rho_0$
do not have more than $E_b \simeq 70 MeV$. If we were to
accept densities closer to $7 \rho_0$ we could hardly get
to $E_b \simeq 150 MeV$. These binding energies may be
too small to allow for the formation and persistence of this
type of strange baryonic matter in the early universe \cite{n}.

\section*{\centering\normalsize \bfseries{8. {Comparison with the LNT Model}}}
\addtocounter{section}{1}
\setcounter{equation}{0}

Since much lower densities, even $1.4 \rho_0$, and larger
binding energies, up to $E_b \simeq 300MeV \!$, have been
found by LNT \cite{lnt},
we want now to discuss the origin of this
difference. 

LNT take the lagrangian in Eq.(\ref{chirallag}) but
without the four and six fermion terms, add to it the lagrangian of
the Walecka model \cite{cw}, a renormalizable lagrangian introducing
two fictitious fields $\phi$ and $V^{\mu}$ coupled to 
baryons, and add the following term coupling $\phi$ to the
pseudo-Goldstone bosons
\begin{equation}
\label{phik}
\half f_{\pi}^2 \Lambda [ Tr {\cal{M}} ( \Sigma - 1) + h.c.]
(b_1 \phi + b_2 \phi^2 + \cdots),
\end{equation}
where $b_1$ and $b_2$ are arbitrary constants and $\phi$
(and $V^{\mu}$) are taken to be chiral singlets.
At a first glance this coupling seems to violate the principle 
of using QCD compatible $\cal{M}$ dependent
terms, obtained by promoting $\cal{M}$ to a field that transforms
under a chiral transformation. This trick apparently would forbid
terms proportional to $Tr \cal{M}$ in the lagrangian. 
It is true that chiral lagrangians do not contain
$\phi$ fields and the rules to construct them do only
apply to mesons and baryons. However, notice that when the
$\phi$ field is heavy, through $\phi$ exchange at tree level
the $b_1$ coupling in Eq.(\ref{phik}) generates an effective
chiral lagrangian coupling proportional to 
$[Tr {\cal{M}}(\Sigma - 1) + .h.c.]^2$, which contains a $Tr \cal{M}$ term.
However, at a second glance one can see that one can start
from the coupling dictated by the trick,
namely $\sim Tr [{\cal{M}} \Sigma + h.c.](b_1 \phi + b_2 \phi^2 + \cdots)$
and by shifting the $\phi$ field to the minimum $\phi_0$ of the
$\phi$ potential for $\Sigma = 1$, namely replacing $\phi$
by $\phi + \phi_0$, one obtains the lagrangian used by LNT, after a
few redefinitions of constants.

LNT use the Walecka model to account for the existence of nuclei,
not included in the chiral lagrangian they study, which does
not have four fermion terms. However it has been shown that 
precisely these terms are equivalent to the Walecka model in bulk
nucleonic matter \cite{gr}. Thus it is not necessary to go out of
chiral lagrangians to account for normal nuclear matter. Besides,
the additional couplings in Eq.(\ref{chirallag}) of 
the Walecka fictitious fields
with mesons are arbitrary. 

We here choose instead to use solely
perturbative chiral lagrangians. We believe this is the main difference.
Although the LNT lagrangian is chirally symmetric it does 
not seem to be obtainable from a perturbative chiral lagrangian.
We will now  compare their solution with the chiral expansion we
used so far. In order to do it, we need to eliminate $\phi$ from the LNT
lagrangian for bulk matter, by using the equation of motion for 
$\phi$ constant,
\begin{equation}
\phi = {g_{\phi} \bar{n}n - f_{\pi}^2 m_K^2 b_1 (1 - \cos \theta)
\over m_{\phi}^2 + 2 f_{\pi}^2 m_K^2 b_2 (1 - \cos \theta)}.
\end{equation}
The LNT lagrangian becomes
\begin{eqnarray}
\label{LNT}
{\cal{L}} &=& \half f_{\pi}^2 \partial_{\mu} \theta
\partial^{\mu} \theta - f_{\pi}^2 m_{K}^2 (1- \cos{\theta}) \nonumber\\
\nonumber\\
&& \, + \bar{n} (i \partial \sl - m_{N})n  +
 a_tm_s(1 - cos \theta) \bar{n} n  \nonumber\\
\nonumber\\
&& \, - {g_{V}^2 \over 2m_{V}^2} (n^{\dagger}n)^2 
+ \half {\left[g_{\phi} \bar{n} n - f_{\pi}^2 m_K^2 
b_1(1 - cos \theta)\right]^2
\over m_{\phi}^2 + 2f_{\pi}^2 m_{K}^2 b_2 (1 - cos \theta)}, 
\end{eqnarray}
where $a_t = 2a_3 + a_2 + a_1 = 2.32 \pm 0.4$, $m_N = 939MeV$ is
the mass of the free neutron (see Eq.(\ref{mns}))
and $g_{\phi}$, $m_{\phi}$, $g_{V}$, $m_{V}$ are respectively 
the couplings to nucleons and the masses of the scalar and vector
Walecka fields.

 Using the
procedure described above (the same used by LNT) one finds the
LNT effective chemical potential and effective mass of the
quasi-particles in bulk matter,
\begin{eqnarray}
\mu^{*} &=& \mu_{B} - {g_v^2 \over m_{v}^2}n^{\dagger}n \\
\label{LNTeffm}
m^{*} &=& m_{N} - a_tm_s(1 - cos \theta) - 
{g_{\phi}^2 \bar n n + g_{\phi} f_{\pi}^2 m_{K}^2 b_1 (1 - cos{\theta})
\over m_{\phi}^2 + 2f_{\pi}^2 m_{K}^2 b_2 (1 - cos{\theta})}.
\end{eqnarray}
These formulas reproduce the results of LNT. 

The main feature of
the LNT model that allows to obtain very low $\rho_n$ and large
$E_b$ is that the value of $a_t$ is $a_t = 8.2 \pm 0.2$ instead
of $a_t = 2.3 \pm 0.2$. We have already seen that large values
of $a_3$ help greatly in reducing the nuclear density of solutions.
This large value of $a_3$ is obtained by LNT, because the exchange
of the $\phi$ particle contributes to pion-kaon scattering so 
that now 
\begin{equation}
\label{a3}
a_3 - {g_{\phi}^2 f_{\pi}^2 b_1 \Lambda \over 2 m_{\phi}^2}
= 1.3 \pm 0.2,
\end{equation}
which for $b_1 = 20 GeV^{-1}$,
the value chosen by LNT, gives $a_3 = 4.2 \pm 0.2$.
Note that if the last terms of Eqs.(\ref{LNT}) and (\ref{LNTeffm})
are expanded in powers of $(1 - \cos \theta)^j$ and the term
in $(1- \cos \theta)$ is summed to the $a_t$ term we obtain
precisely
Eq.(\ref{a3}) (one needs to refer to the original lagrangian
to see that only $a_3$ and not $a_1$ and $a_2$ are corrected).
Let us write the terms in Eq.(\ref{LNT}) that depend solely on $\theta$
(not on $n$). These will be enough for our argument.
\begin{eqnarray}
\label{utheta}
{\cal U}_{\theta}^{LNT} =  f_{\pi}^2 m_{K}^2 (1- \cos{\theta}) -
\half {f_{\pi}^4 m_K^4 {b_1}^2(1 - cos \theta)^2
\over m_{\phi}^2 + 2f_{\pi}^2 m_{K}^2 b_2 (1 - cos \theta)}.
\end{eqnarray}

To rewrite these terms in a form similar to chiral perturbation theory
we use the Taylor expansion
\begin{equation}
( 1 + \beta(1 - cos{\theta}))^{-1} = (1 + \beta)^{-1}
\left( 1 + \sum_{j=1}^{\infty} { \left( {\beta \over 1 + \beta} \right)^j
\cos^j \theta} \right),
\end{equation}
where $\beta = 2 f_{\pi}^2 b_2 (m_{K} / m_{\phi})^2 \simeq 1.66$,
to expand the denominator of Eq.(\ref{utheta}) which
becomes,
\begin{eqnarray}
\label{uu}
{\cal U}_{\theta}^{LNT} &=& - f_{\pi}^2 m_K^2 \left(
1 -  \alpha { 2 + \beta \over 1 + \beta}
\right) \cos \theta  \nonumber\\
&& \,- f_{\pi}^2 m_K^2 {\alpha \over (1 + \beta)^2} \sum_{j=0}^{\infty} {
\left( {\beta \over 1 + \beta} \right)^j {\cos}^{j+2} \theta}, 
\end{eqnarray}
where
\begin{eqnarray}
\hspace{-1in}
\alpha &=& \half { f_{\pi}^2 m_K^2{b_1}^2 \over (1 + \beta) m_{\phi}^2}
\simeq 0.62,
\end{eqnarray}
once the constant term has been dropped.

In chiral perturbation theory the terms of
${\cal{U}}_{\theta}$ of power $j$ in $cos{\theta}$ are proportional
to $(m_s/\Lambda)^j= 0.24 ^j$, so naively we would expect each term to
be 0.24 of the previous term.  Moreover Eq.(\ref{uchiral})
 shows that the first order
contribution to the kaon mass accounts for 0.94 of the total mass. 
 Let us isolate the kaon mass term in Eq.(\ref{uu}) by
expanding ${\cos} \theta$ to order
$\theta^2 = 2{\langle K^{0}\rangle}^2/ f_{\pi}^2$ , 
\begin{eqnarray}
\label{utheta2} 
{\cal U}_{\theta}^{LNT}\!\! &=& \!\! m_K^2 {\langle K^{0}\rangle}^2 \left[
1 - \alpha { 2 + \beta \over 1 + \beta} 
+ {\alpha \over (1 + \beta)^2} \sum_{j=0}^{\infty} {
\left( {\beta \over 1 + \beta} \right)^j (j + 2)} \right]+ \cdots
\end{eqnarray}
The terms in the parenthesis of Eq.(\ref{utheta2}) sum to unity as we expect
from Eq.(\ref{utheta}) (where the only $\theta^2$ term is
$m_K^2 \langle K^0 \rangle^2$). 
However, writing out the first
few terms explicitly in Eq.(\ref{utheta2}), 
\begin{eqnarray}
{\cal U}_{\theta}^{LNT} &\simeq& m_K^2 {\langle K^{0}\rangle}^2
( 0.15 + 0.18 + 0.16 + 0.14 + \cdots)+ \cdots 
\end{eqnarray}
we see that the series converges very slowly
in comparison to the  perturbative expansion in chiral lagrangians
shown in Eq.(\ref{uchiral}).

Another way of showing the same difference between the LNT models and ours
is by 
 obtaining the LNT coefficient $b_1$ in terms of the
chiral expansion coefficient J.
This is most easily done by expanding ${\cal U}_{\theta}$ 
in powers of $(1- \cos{\theta})$ instead of powers of $\cos{\theta}$ \begin{equation}
\label{u1}
{\cal U}_{\theta} = (f_{\pi}^2 \tilde{m}_{K}^2 + 
 2 J f_{\pi}^2 {\tilde{m}_{K}^4 \over \Lambda^2})(1- \cos{\theta})
 - J f_{\pi}^2 {\tilde{m}_{K}^4 \over \Lambda^2} (1 - cos \theta)^2 + \cdots .
\end{equation}

Writing ${\cal U}_{\theta}$ in this way it becomes obvious
 that the higher order term for $J$ positive lowers the potential energy of
the condensate. Notice the factor multiplying $(1- \cos{\theta})$ in the
first term, is 
$f_{\pi}^2 m_K^2$ as shown in Eq.(\ref{uchiral}).

Referring back to Eq.(\ref{utheta}), we also expand $\cal L$ in
terms of $(1 - \cos \theta)$ (only for $\theta < 1.16$, so
that $2 f_{\pi}^2 m_K^2 b_2 (1 - cos{\theta}) / m_{\phi}^2 < 1$) and obtain
\begin{eqnarray}
\label{u2}
{\cal U}_{\theta}^{LNT} &=&  f_{\pi}^2 m_{K}^2 (1- \cos{\theta}) -
{f_{\pi}^4 m_{K}^4 b_1^2 \over 2 m_{\phi}^2} (1 - \cos \theta)^2 \nonumber\\
\nonumber\\
&& \, - {m_{\phi}^2 \over 8} \left( {b_1 \over b_2} \right)^2
 \sum_{j=3}^{\infty} {(-1)^j \left({2 f_{\pi}^2 b_2 m_{K}^2 \over
m_{\phi}^2} \right)^j (1 - \cos \theta)^{j}.}
\end{eqnarray}
Comparing the coefficients of the terms proportional to 
$(1 - \cos \theta)^2$ in Eqs.(\ref{u1})
and (\ref{u2}), from the bound $J < 0.14$ we obtained above
 we get $b_1 < 2.7{GeV}^{-1}$,
which is much smaller than $20{GeV}^{-1}$, the value
used by LNT.
Thus, even if the LNT lagrangian is chirally symmetric it does 
not seem to be obtainable from a perturbative chiral lagrangian.

\begin{center}
\section*{\normalsize \bfseries{9. {Conclusions}}}
\end{center}
\addtocounter{section}{1}
\setcounter{equation}{0}

The main results of this paper are given in Fig.3 where we
show the lowest baryonic number densities $\rho_n$ corresponding to
 infinite solutions with  a given binding energy $E_b$, that
we obtain with a chiral effective lagrangian, as explained
in section 7. Actually only
the region of $\rho_n < 7\rho_0$ is consistent with our approach.
In section 5 we found that the lowest order \suthree chiral
lagrangian, including four fermion terms, could not 
yield solutions with densities smaller than $4.5 \rho_0$ 
(see Fig.3) and at these densities the fermion bilinear
expansion parameter $\bar{B} \Gamma_{\alpha} B / f_{\pi}^2 \Lambda
\simeq \rho_n / 7\rho_0$ is large and higher order terms with
more baryons should be considered. We analyze them in
section 6 and find that they do not help to lower
$(\rho_n)_{min}$ for a given binding energy. In section
7 we show how much higher order condensate dependent terms
help, when their coefficients are chosen to respect
the perturbative expansion. We find that solutions with
$\rho_n < 4\rho_0 (7\rho_0)$ have at most
only $50 MeV (150MeV)$ of binding energy. We believe
the perturbative chiral lagrangian cannot reasonably 
do any better.
The  densities we obtain are entirely compatible with the
onset of a K-condensate at densities 2--4 $\rho_0$.

 The binding
energies we find may be too low to allow for the formation and
persistence of this type of strange baryonic matter in the 
Universe at temperatures of order $100 MeV$ \cite{n}.
However, we have used only a $K^0$--condensate and neutrons,
the lightest baryons in such a condensate. We may not 
exclude that a less restrictive ansatz may provide solutions
with larger binding energies (a factor of 3 or 4 may suffice
\cite{n}) that may appear and persist in the early Universe.

\vskip0.3in
\par\noindent
{\bf Acknowledgements}
\vskip0.1in

We thank B. Lynn for his contribution to the beginning stages of
this work and A. Nelson for discussions about her earlier work
on this subject.
The authors are partially supported by  the US Department of Energy
under grant DE-FG03-91ER40662 Task C.

\pagebreak

\appendix
\setcounter{equation}{0}
\section*{\centering\normalsize \bfseries {Appendix A}}
\setcounter{section}{1}
These are complete equations with all higher order terms included in this
paper for the effective neutron chemical potential, effective neutron
mass, pressure and energy density, that we label with the sub index
``$_{TOTAL}$''. See Eqs.(\ref{density}),
(\ref{scalarden}) and (\ref{vectorden}) for $(n^{\dagger} n)$, $(\bar{n}n)$
and $(\bar{n} \vec{\gamma}\cdot \vec{k} n)$, respectively.
\begin{eqnarray}
\nonumber\\
\mu^*_{\scriptscriptstyle{TOTAL}} &=& 
 \left[ \mu_B - {C_V^2 \over f_{\pi}^2}
(n^{\dagger} n) + {2 G_2 \over f_{\pi}^4 \Lambda} (\bar{n}n)
(n^{\dagger}n)\right.\nonumber\\
&& \hbox{\hspace{1in}} \left. - {I_V m_s \over f_{\pi}^2 \Lambda}
(1 - \cos \theta) (n^{\dagger}n) \right] \left[ 
1 + {H_1 \over f_{\pi}^2 \Lambda}(\bar{n} n) \right]^{-1} \nonumber\\
&& \\
&& \nonumber\\
 m^*_{\scriptscriptstyle{TOTAL}} &=& 
 \left[ m^*_N - {C_S^2 \over f_{\pi}^2}
(n^{\dagger} n) -  {G_1 \over f_{\pi}^4 \Lambda} (\bar{n}n)^2
- {G_2 \over f_{\pi}^4 \Lambda} (n^{\dagger}n)^2 \right. \nonumber\\
&& \hbox{\hspace{.1in}} \left.+ {H_1 \over f_{\pi}^2 \Lambda}
 (\bar{n} \vec{\gamma} \cdot \vec{k} n)
- {I_S m_s \over f_{\pi}^2 \Lambda}
(1 - \cos \theta)(\bar{n} n)\right] \left[ 1 + {H_1 \over f_{\pi}^2 \Lambda}
(\bar{n} n) \right]^{-1} \nonumber\\
&& \\
&&  \nonumber\\
P_n^{\scriptscriptstyle{TOTAL}} &=&
{1 \over 3} (1 + {4H_1 \over f_{\pi}^2 \Lambda} \bar{n}n)
(\bar{n}  \vec{\gamma} \cdot \vec{k} n) +
 {C_V^2 \over 2 f_{\pi}^2} -  {C_S^2 \over 2 f_{\pi}^2}
- {2 G_1 \over 3 f_{\pi}^4 \Lambda} (\bar{n}n)^3 \nonumber\\
&& \, - {2 G_2 \over f_{\pi}^4 \Lambda} (\bar{n}n)(n^{\dagger}n)^2
+ {I_V m_s \over 2 f_{\pi}^2 \Lambda}(1 - \cos \theta) (n^{\dagger}n)
- {I_S m_s \over 2 f_{\pi}^2 \Lambda}(1 - \cos \theta)(\bar{n} n) \nonumber\\
&& \\
&& \nonumber\\
&& \nonumber\\
\varepsilon_{\scriptscriptstyle{TOTAL}} &=&
 2 \int_{}^{k_f}{{d^{3}k \over
 (2 \pi)^{3}}(k^2 + m^{*2})^{\half}} +  {H_1 m^* \over f_{\pi}^2 \Lambda}
(\bar{n}n)^2
+ {C_{V}^2 \over 2f_{\pi}^2}(n^{\dagger}n)^2 
+ {C_{S}^2 \over 2f_{\pi}^2}( \bar{n}n)^2 \nonumber\\
&& \, +  {2 G_1 \over 3 f_{\pi}^4 \Lambda} (\bar{n}n)^3 
+ {I_V m_s \over 2 f_{\pi}^2 \Lambda}(1 - \cos \theta) (n^{\dagger}n)
+ {I_S m_s \over 2 f_{\pi}^2 \Lambda}(1 - \cos \theta)(\bar{n} n)
+ {\cal{U}}_{\theta} \nonumber\\
&& \\
&& \nonumber\\
&& \nonumber\\
{\cal{U}}_{\theta} &=& m_s \Lambda f_{\pi}^2 (1 - \cos \theta)
+  J f_{\pi}^2 m_s^2
(1 - \cos^2{\theta})
\end{eqnarray}
\normalsize

\pagebreak

\pagebreak

\begin{table}[p]
\centerline{
\begin{tabular}{||c|c|c|r|c|c|c||}                          \hline
{$(C_S^2)_{max}$} & $(C_V^2)_{max}$ &
\multicolumn{2}{c|}{$n \geq 3$} & $(\rho_n)_{min} / \rho_0$ &
$m^* (MeV)$ & $\theta_0$  \\ \hline
$0.62$ & $0.24$ & --- & --- & $6.6$ & $220$ & $1.4$           \\ \hline
$0.00$  & $0.17$ & \raisebox{-.1in}[1.5ex][.75ex]{$G_1$} 
                & $0.65$ & $8.1$ & $180$ & $1.7$        \\ \cline{4-7} \cline{1-2}
$0.77$ & $0.21$ &                   
                & $-1.00$ & $7.1$ & $200$ & $1.6$               \\ \hline
$0.61$ & $0.27$ & \raisebox{-.1in}[1.5ex][.75ex]{$G_2$} 
                & $0.04$ & $6.6$ & $230$ & $1.3$        \\ \cline{4-7} \cline{1-2}
$0.52$ & $0.01$ &                   
                & $-0.25$ & $6.9$ & $210$ & $2.0$                 \\ \hline
$0.58$ & $0.05$ & \raisebox{-.1in}[1.5ex][.75ex]{$H_1$} 
                & $0.50$ & $7.8$ & $190$ & $1.9$        \\ \cline{4-7} \cline{1-2}
$0.64$ & $0.28$ &                   
                & $-0.10$ & $6.3$ & $240$ & $1.3$            \\ \hline
$0.34$ & $0.02$ & \raisebox{-.1in}[1.5ex][.75ex]{$I_V$} 
                & $0.80$ & $7.3$ & $342$ & $1.3$        \\ \cline{4-7} \cline{1-2}
$0.89$ & $0.48$ &                   
                & $-0.90$ & $7.6$ & $73$ & $1.8$                \\ \hline
$0.72$ & $0.36$ & \raisebox{-.1in}[1.5ex][.75ex]{$I_S$} 
                & $3.00$ & $5.3$ & $210$ & $1.2$        \\ \cline{4-7} \cline{1-2}
$0.58$ & $0.19$ &                   
                & $-1.00$ & $7.4$ & $230$ & $1.6$               \\ \hline
$0.66$ & $0.26$ & $J$ & $0.14$ & $6.4$ & $210$ & $1.4$         \\ \hline
\end{tabular}}

\caption{This table shows the values of $(C_S^2)$, $(C_V^2)$,
the one higher order
parameter chosen to be non-zero, the baryon number density,  the
effective mass $m^*$ of the quasi particles and  the
condensate $\theta_0$, with which the baryon number density is minimum
for a fixed binding energy $E_b=-39MeV$ and parameter $a_3=1.3$.
This corresponds to the  maximum values of $(C_S^2)$ and $(C_V^2)$ for 
which solutions exist, after fixing the higher order parameter to the largest
positive and negative physically acceptable values. The first entry 
is the solution in Figs.1 and 2. The strangeness/baryon for these solutions
is $(1 - \cos{\theta_0})$ which lies in the range $0.64$ -- $1.4$.}
\end{table}

\pagebreak
\begin{figure}[p]
\centerline{\epsffile{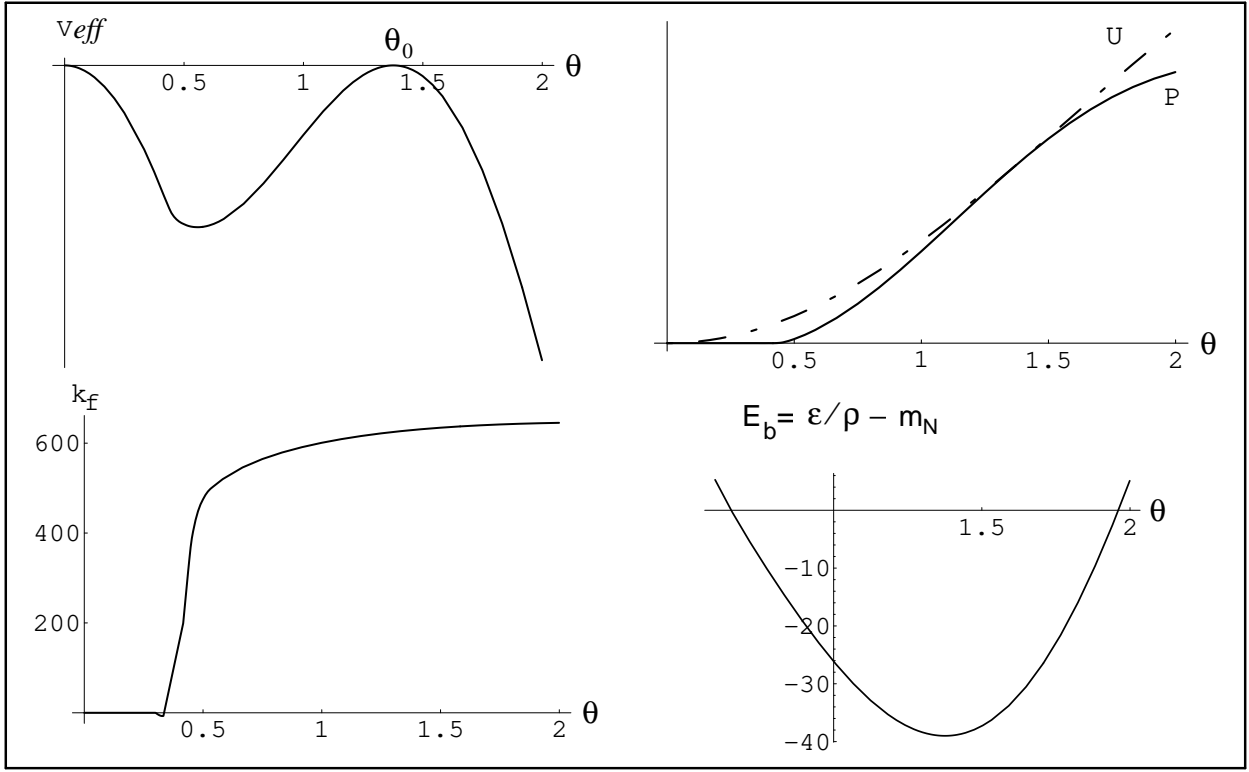}}
\caption{An infinite  solution with binding energy 
$E_b = -39 MeV$, number density $\rho_n = 6.6 \rho_0$, and
parameters $C_V^2 = .24\ \mbox{and}\ C_S^2 = .62$ (and no higher order
terms). 
This is a minimum density solution where
$\rho = \rho_{min}$, $C_V^2 = (C_V^2)_{max}$, and
$C_S^2 = (C_S^2)_{max}$. It corresponds to the first entry in the Table 1.}
\end{figure}

\pagebreak
\begin{figure}[p]
\centerline{\epsffile{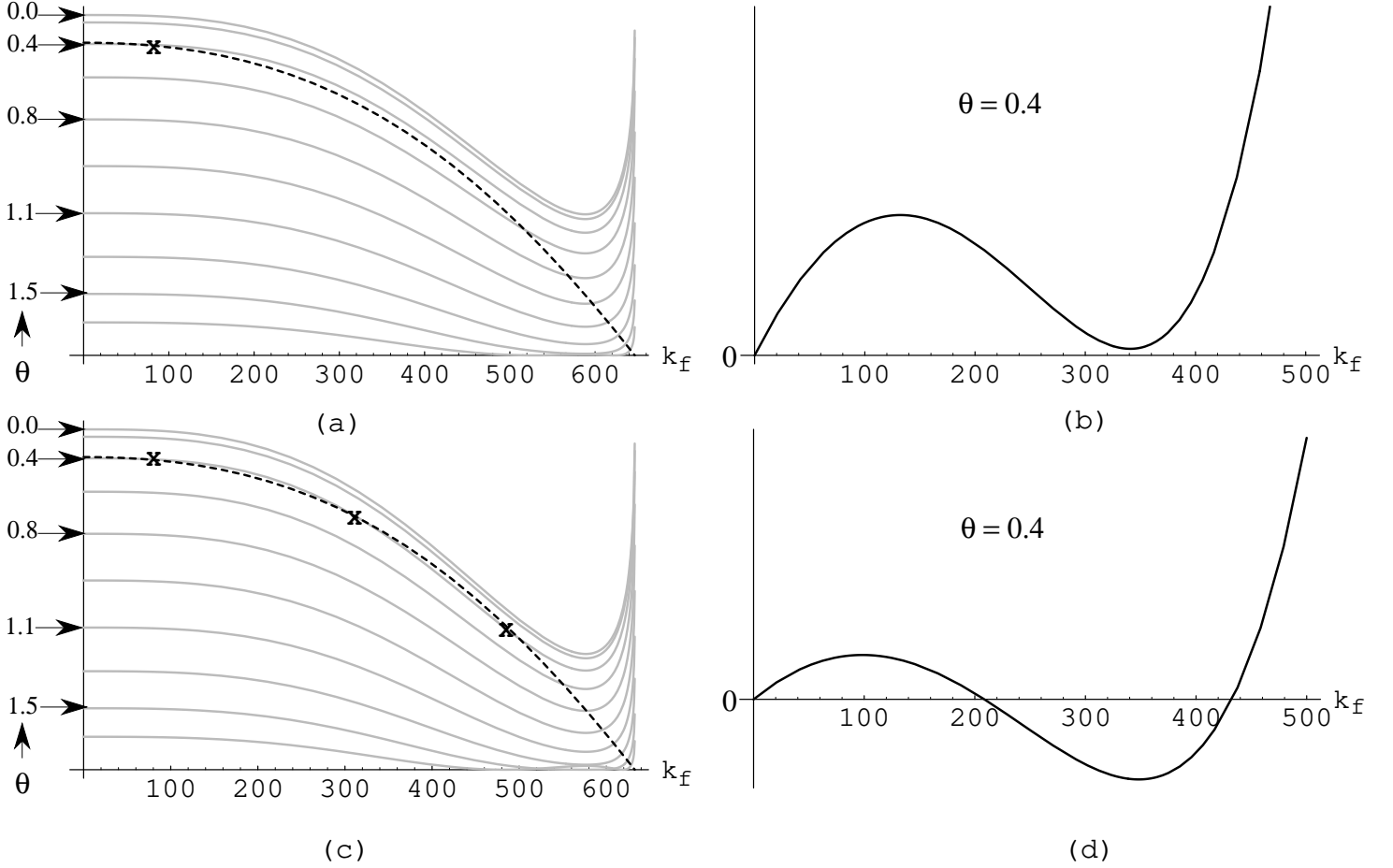}}
\caption{(a) The l.h.s. and r.h.s of Eq.(\ref{disper}) for
the solution shown in Fig.1 are plotted here. The dark
dashed line corresponds to the l.h.s of Eq.(\ref{disper}),
and the gray lines to the r.h.s side for the 
indicated values of $\theta$.
The intersection of these curves determines the numerical solution for
$k_f( \theta)$. (b) This is a plot of the function
$2 m^* (dm^*/dk_f) - 2 \mu^* (d\mu^*/dk_f) - 2k_f$, which is the
l.h.s. of Eq.(\ref{dercond}), for $\theta = 0.4$. This shows 
why the solution in Fig.1 has  minimum density, since for
any higher values of $C_V^2$ and $C_S^2$ the function shown here
would have negative values and, hence, the solution would be
unphysical.
(c),(d) Show the same as (a) and (b) respectively, but  for
an unphysical solution with the same
binding energy, with parameters $
C_V^2 = .27\ \mbox{and}\ C_S^2 = .77$, larger than
$(C_V^2)_{max} = .24$ and $(C_S^2)_{max} = .62$ corresponding to the
case in (a) and (b).}
\end{figure}

\pagebreak
\begin{figure}[htb]
\centerline{\epsffile{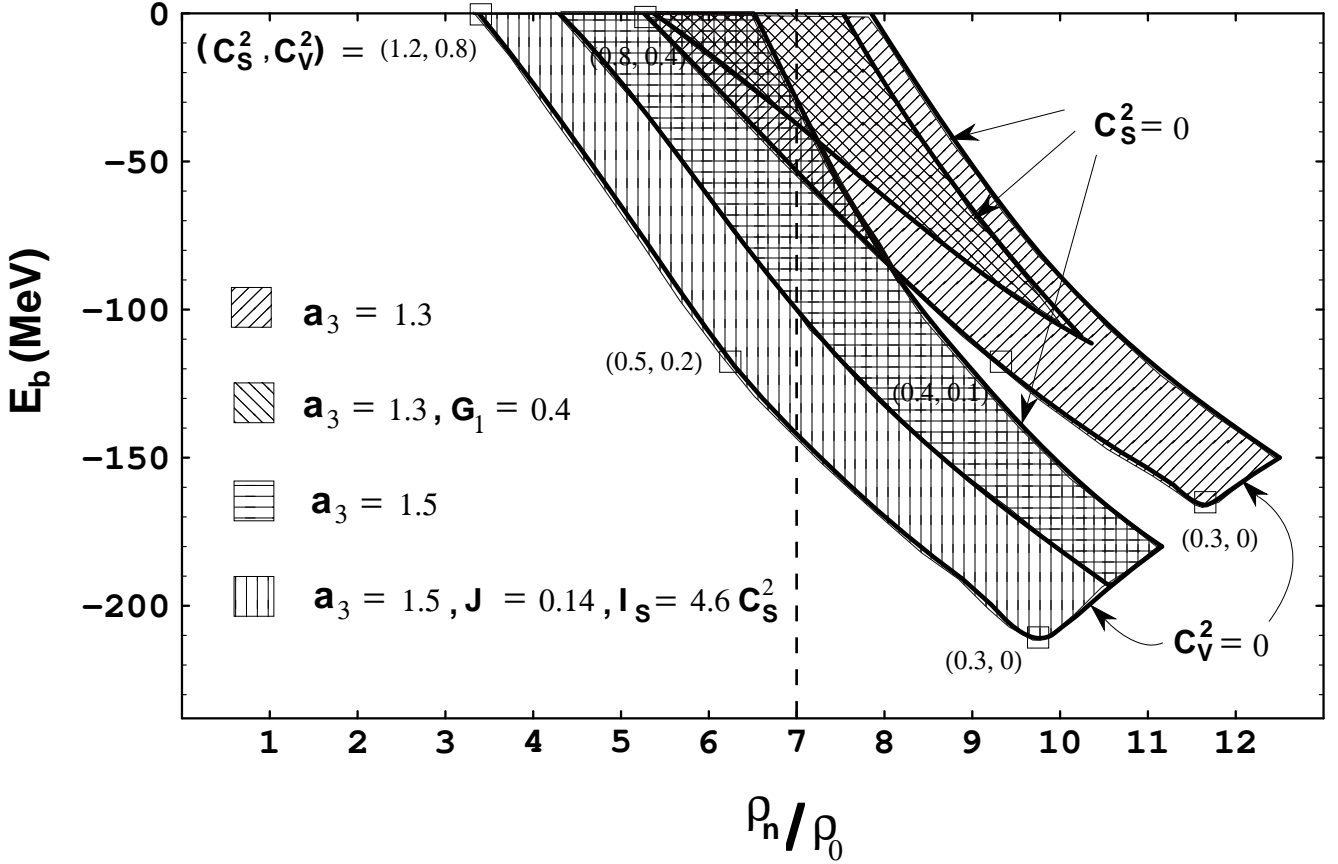}}
\caption{Here we show the regions of physical solutions 
in ($E_b$, $\rho_n$) space corresponding to the indicated 
parameters. We indicate only the values of $a_3$ and 
the higher order parameters taken to be non-zero. We then find the
values of $C^2_S$ and $C^2_V$ at the left and right boundaries of each
band with the procedure described in the text.  $C^2_S$ and $C^2_V$
are maximum at the left boundaries and $C^2_S=0$ at the right ones.
The lower boundary corresponds to $C^2_V=0$. The values of 
($C^2_S$,$C^2_V$) are shown at some points.
 We have chosen the most favorable values of $a_3$ and the
 higher order parameters to obtain the lowest baryonic number densities 
 $\rho_n$ for infinite solutions with  a given binding energy
 $E_b$, using a chiral effective lagrangian. Actually only
the region of $\rho_n < 7\rho_0$ is consistent with our approach
(the perturbative expansion is lost at higher densities), so the rest
of the figure may be indicative but is not believable.}
\end{figure}
\end{document}